\documentclass[prl, twocolumn, reprint]{revtex4-2}
\usepackage{amsmath}
\usepackage{amsfonts}
\usepackage{amssymb}
\usepackage{graphicx}
\usepackage{bbm}
\usepackage{mathrsfs} 
\usepackage{natbib}
\usepackage{color}
\usepackage{mathtools}
\usepackage{physics}  

\usepackage[colorlinks=true, 
            linkcolor=blue,
            citecolor=blue,
            urlcolor=blue,
            pdfstartview=FitH,
            bookmarks=true,
            bookmarksopen=true]{hyperref}
\begin{document}

\title{Unsupervised Topological Phase Discovery in Periodically Driven Systems via Floquet-Bloch State}

\author{Chen-Yang Wang}
\affiliation{School of Physics Science and Engineering, Tongji University, Shanghai 200092, China}

\author{Jing-Ping Xu}
\affiliation{School of Physics Science and Engineering, Tongji University, Shanghai 200092, China}

\author{Ce Wang}
\email{cewang@tongji.edu.cn}
\affiliation{School of Physics Science and Engineering, Tongji University, Shanghai 200092, China}

\author{Ya-Ping Yang}
\affiliation{School of Physics Science and Engineering, Tongji University, Shanghai 200092, China}

\begin{abstract}

Floquet engineering offers an unparalleled platform for realizing novel non-equilibrium topological phases. However, the unique structure of Floquet systems, which includes multiple quasienergy gaps, poses a significant challenge to classification using conventional analytical methods. We propose a novel unsupervised machine learning framework that employs a kernel defined in momentum-time ($\boldsymbol{k},t$) space, constructed directly from Floquet-Bloch eigenstates. This approach is intrinsically data-driven and requires no prior knowledge of the underlying topological invariants, providing a fundamental advantage over prior methods that rely on abstract concepts like the micromotion operator or homotopic transformations. Crucially, this work successfully reveals the intrinsic topological characteristics encoded within the Floquet eigenstates themselves. We demonstrate that our method robustly and simultaneously identifies the topological invariants associated with both the $0$-gap and the $\pi$-gap across various symmetry classes (1D AIII, 1D D, and 2D A), establishing a robust methodology for the systematic classification and discovery of complex non-equilibrium topological matter.

\end{abstract}
\maketitle

\textit{Introduction}-Topological condensed matter physics has achieved landmark progress in recent decades, with its core lying in the robust edge states protected by topological invariants and bulk energy gaps~\cite{topology1,topology2,topology3,topology4,topology5,topology6,topology8,topology9,topologyhaldane,topologyssh,topologyssh1,topologyssh2}. While traditional classification schemes primarily focused on equilibrium systems described by static (time-independent) Hamiltonians, the rapid development of experimental techniques has spurred interest in extending topological concepts to non-equilibrium states. This is particularly exemplified by time-periodically driven systems, known as Floquet systems~\cite{floquet1,floquet2,floquet3,floquet4,floquet5,floquet6,floquet7,floqueta,floquetbdi1,floquetbdi2,floquetd1,floquetd2,floquetsingularities,floquetten,floquetd3,floquetbdi3}.

Floquet engineering provides a powerful paradigm for the active manipulation and design of quantum matter. By applying periodic external driving, the system’s dynamics can be described by a time-independent effective Hamiltonian. The spectrum of this Hamiltonian is composed of quasi-energies, the inherent periodicity of which brings a rich topological structure far beyond that of static Hamiltonians. A notable feature is the emergence of non-trivial topological edge states even when the effective (time-averaged) Hamiltonian is trivial~\cite{floqueta}.

The unparalleled designability of Floquet systems, achieved by tuning periodic driving parameters, offers unprecedented possibilities for realizing novel non-equilibrium topological phases. However, this richness brings classification challenges that exceed those of static systems. Specifically, the quasi-energy spectrum's periodicity introduces multiple independent and interrelated topological gaps, and the highly nonlinear dependence of the phase diagram on driving parameters makes systematic exploration using traditional analytical or numerical methods inherently difficult~\cite{floquetbdi2,floquetd3}.

This is where unsupervised learning provides a crucial advantage. The task of identifying phase transitions and classifying phases of matter using unsupervised methods has become a rapidly developing and vital area of research~\cite{Un1,un2,un3,un4,un5,un6,un7,Carrasquilla2017,Wang2018,PhysRevB.96.144432,diffusion1,diffusion2,diffusion3,diffusion4,diffusion5,diffusion6,diffusion7,diffusionlongyang,diffusionfloquet}. Early efforts in this field used techniques like Principal Component Analysis (PCA) to successfully identify phase transitions in classical statistical models~\cite{Carrasquilla2017,Wang2018,PhysRevB.96.144432}. Subsequently, advanced manifold learning methods, such as diffusion maps, demonstrated success in clustering topological phases based on state information~\cite{diffusion1,diffusion2,diffusion3,diffusion4,diffusion5,diffusion6,diffusion7,diffusionlongyang,diffusionfloquet}. A major, more recent advancement involves using kernel methods based on projection operators for the non-iterative clustering of gapped quantum systems ~\cite{diffusionlongyang}.

Inspired by these powerful, emerging approaches, we extend the efficient kernel-based unsupervised learning framework to address the complexities of Floquet topological systems. We introduce a novel kernel defined directly in the momentum-time ($\boldsymbol{k},t$) space. This kernel is designed to effectively quantify the topological similarity between Floquet Hamiltonians by evaluating the geometric structure of their eigenstates. A prior unsupervised approach for Floquet topological insulators relied on the abstract and computationally involved micromotion operator and specific homotopic transformations, often demanding prior knowledge of the system~\cite{diffusionfloquet}. \textbf{In sharp contrast}, our $\boldsymbol{k},t$-space kernel utilizes information directly from the Floquet eigenstates, offering a computationally efficient and fundamental data-driven methodology that requires no reliance on the micromotion operator, homotopic transformations, or any prior topological invariant knowledge.
\begin{figure*}[tbh!] 
	\centering
    	\includegraphics[angle=0,width=18.5cm]{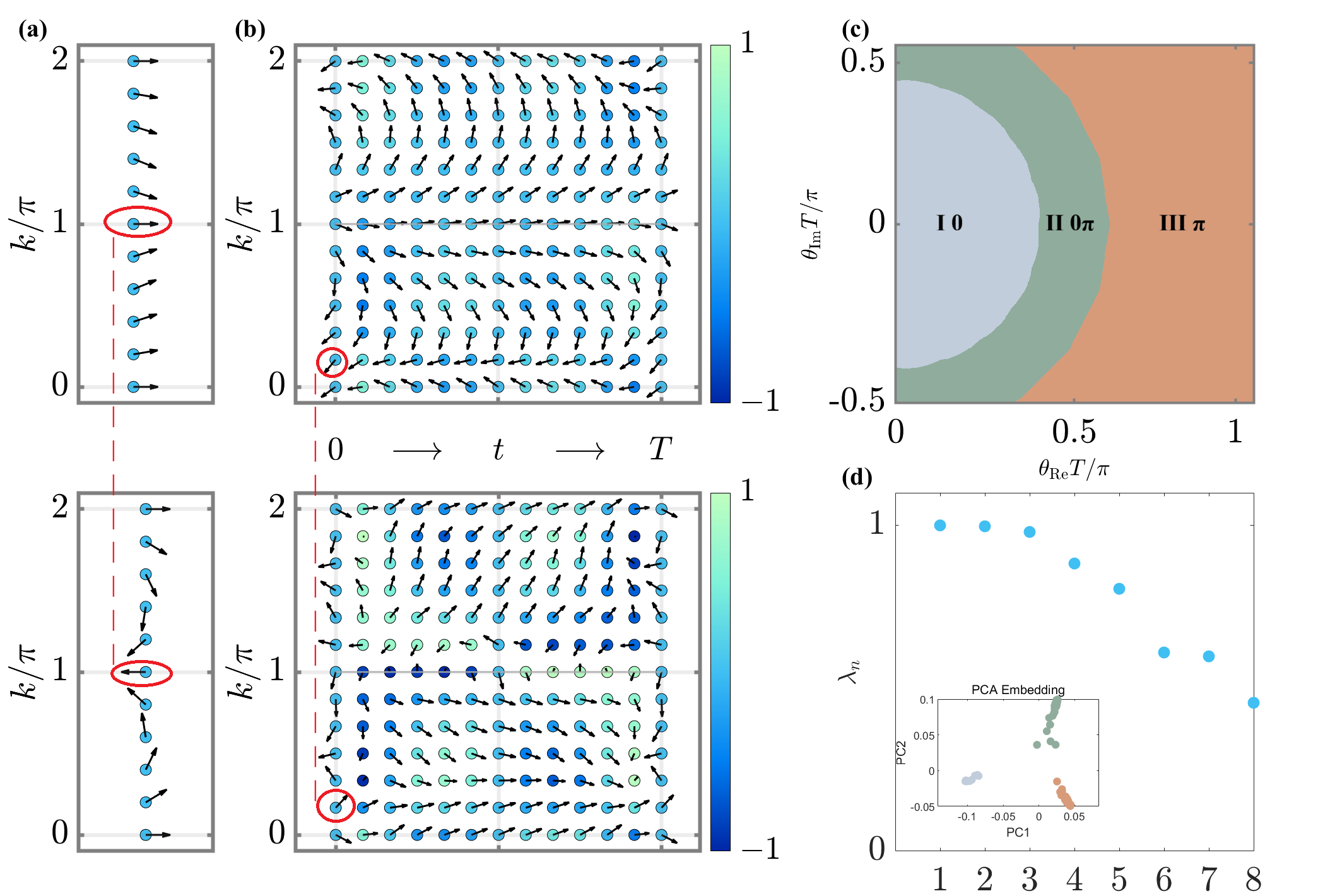}

	\caption{(a) Two static topological systems possessing distinct topological numbers must necessarily exhibit band inversion points. The top and bottom figures illustrate the Bloch vector representations for the topologically trivial and non-trivial phases, respectively, in the SSH model. (b) The Bloch vector representations for the \textit{flattened Floquet operator} for the modulated Floquet SSH model. The top configuration is sampled from the $0$-phase ($\theta_{\mathrm{Re}}=0.2\pi/T,\theta_{\mathrm{Im}}=0.2\pi/T$) and the bottom configuration is sampled from the $\pi$-phase ($\theta_{\mathrm{Re}}=0.6\pi/T,\theta_{\mathrm{Im}}=0.2\pi/T$). The colorbar describes the $z$-component of the Bloch vector.  (c) The phase diagram is derived using the unsupervised clustering algorithm. Comparing the result with the theoretical topological numbers, we observe that Phase I has $W_0 = 1, W_\pi = 0$, Phase II has $W_0 = 1, W_\pi = 1$ and Phase III has $W_0 = 0, W_\pi = 1$. (d) The eigenvalues output by the diffusion map algorithm, where three eigenvalues close to 1 indicate that the data has been partitioned into three clusters. The inset shows the two-dimensional projection of the dataset in the principal component space after sufficient diffusion. }
	\label{fig1}
\end{figure*}

We demonstrate that our kernel enables the robust, simultaneous identification of the topological invariants of the phases associated with both the $0$-gap and the $\pi$-gap across different symmetry classes (AIII, D, and A). Our results establish a clear data-driven approach for identifying Floquet phases. We further show that the Floquet eigenstates in $\boldsymbol{k},t$-space inherently encode the topological information for both gaps simultaneously. This framework can be readily extended to classify other complex Floquet topological systems.

\textit{Method }- First, we review the core concept of unsupervised topological classification for static bands: For any two topologically distinct bands, there exists a momentum $\boldsymbol{k}$ where the sum of their corresponding \textit{flattened Hamiltonians} yields at least one zero eigenvalue~\cite{diffusionlongyang}. This is schematically illustrated for the static Su–Schrieffer–Heeger (SSH) model~\cite{topologyssh,topologyssh1,topologyssh2} in Fig.~\ref{fig1}(a), where the flatten Hamiltonian corresponds to the Bloch vector. Crucially, configurations with distinct winding numbers exhibit antiparallel Bloch vectors at certain momentum points. Building on this observation, one can design a kernel function that assigns larger values when the determinant of the summation vanishes at a given $\boldsymbol{k}$-point, and smaller values otherwise. The topological classification of data sets is subsequently achieved by integrating this kernel with an appropriate unsupervised clustering algorithm~\cite{diffusionlongyang}.

We now extend this idea to Floquet systems. 
 The eigenmode of a periodically driven lattice can be represented by the \textbf{Floquet-Bloch state}~\cite{floquet6,floquetten}$\left|\phi_{n}(\boldsymbol{k},t)\right\rangle$ with 
\begin{equation}
    \left( H(\boldsymbol{k},t)-i\frac{\partial }{\partial t}\right)\left|\phi_{n}(\boldsymbol{k},t)\right\rangle=\epsilon_n(\boldsymbol{k})\left|\phi_{n}(\boldsymbol{k},t)\right\rangle,
\end{equation}
where $H(\boldsymbol{k},t) = H(\boldsymbol{k},t+T)$ is the Hamiltonian, $n$ is the band index and $\varepsilon_n (\boldsymbol{k})\in [-\pi/T,\pi/T ]$ is the quasi-energy of $n$-th band (out of a total of $N_{\text{b}}$ bands). We define the \textit{flattened Floquet operator} (FFO) as
\begin{equation}
    Q(\boldsymbol{k},t) =1-2\sum_{n\in occ}\left|\phi_{n}(\boldsymbol{k},t)\right\rangle \left\langle\phi_{n}(\boldsymbol{k},t)\right|,
\end{equation}
where $occ$ means the occupied Floquet bands. Crucially, we demonstrate through various examples that the topology of Floquet-Bloch states is profoundly linked to specific-gap Floquet topological invariants. Consequently, the FFO plays an analogous role to the flattened Hamiltonian in static systems: For any two topologically distinct Floquet bands, there exists a spacetime point 
$(\boldsymbol{k},t)$ where the sum of their corresponding FFOs yields at least one zero eigenvalue. As show in Fig.~\ref{fig1}(a,b), we have encircled the spin-flip points in red. At these spin-flip points, one eigenvalue of FFO becomes zero.

To achieve unsupervised classification of Floquet Hamiltonians randomly sampled in the parameter space, we define the kernel matrix:
\begin{equation}
K_{ij} = \prod_{\substack{\boldsymbol{k} \in \mathrm{BZ} \\ t \in [0,T)}} 
    \Biggl( 1 - \exp\Biggl[ -\,
    \biggl| \frac{\det\left( Q^{(i)}(\boldsymbol{k},t) + Q^{(j)}(\boldsymbol{k},t) \right)}{\epsilon} \biggr|^2 \Biggr] \Biggr),
\end{equation}
where superscripts $i,j$ index distinct parameter configurations and $\epsilon \to 0^+$ is a sufficiently small positive constant that ensures $K_{ij}$ can be
approximately treated as a binary step function.  This matrix quantifies pairwise topological similarities among $N$ sampled Hamiltonians, satisfying $K_{ij} \to 0$ for topologically inequivalent pairs.

To classify the dataset into topologically distinct clusters, we implement the \textit{diffusion map} algorithm by constructing the probability transition matrix $P_{ij} = K_{ij} / \sum_{j'} K_{ij'}$. The diffusion distance~\cite{diffusion1,diffusion2,diffusion3,diffusion4,diffusion5,diffusion6,diffusion7,diffusionlongyang,diffusionfloquet} between samples $i$ and $j$ after $\ell$ steps is given by:
\begin{equation}
d^{(\ell)}_{ij} = \sqrt{ \sum_{n=0}^{N-1} \lambda_{n}^{2\ell} \left[ (v_{n})_i - (v_{n})_j \right]^2 }
\end{equation}
where $v_n$ is the $n$th right eigenvector of the normalized transition matrix $\hat{P}$ with corresponding eigenvalue $\lambda_n$ ($n=0,1,\dots,N-1$), ordered such that $1 = \lambda_0 \geq \lambda_1 \geq \cdots \geq \lambda_{N-1} \geq 0$. Under prolonged diffusion ($\ell \gg 1$), the dominant contributions arise from the eigenvectors $v_n$ with $\lambda_n \approx 1$. These components encode the essential topological features of the diffusion manifold, forming a low-dimensional representation of the original FFO data. Topological classification is then derived by clustering this reduced representation.

Next, we apply this framework to Floquet systems across different symmetry classes.

\textit{AIII class in one-dimension} - As a concrete example, we consider the modulated Floquet SSH model in symmetry class AIII, described by the Hamiltonian~\cite{floquetbdi1,floquetbdi2,floquetbdi3}:
\begin{equation}
H(k,t) = \theta_{\text{Re}} \sigma_x + \theta_{\text{Im}} \sigma_y + \gamma(t) \left[ \cos(k) \sigma_x + \sin(k) \sigma_y \right],
\end{equation}
where $\theta_{\text{Re}}$ and $\theta_{\text{Im}}$ represent the real and imaginary parts of the intracell coupling, $\gamma(t)=0.6\pi/T+2g\cos(\omega t)$ is a periodically oscillating even function that models the intercell coupling, and $\sigma_x, \sigma_y, \sigma_z$ are the Pauli matrices. A key aspect of this model is the inclusion of an imaginary intercell coupling, which explicitly breaks time-reversal and particle-hole symmetries. This symmetry breaking removes high-symmetry points along the momentum direction, thereby substantially enriching the diversity and generality of the dataset. The system retains chiral symmetry~\cite{floquetten}, defined by $\mathcal{S}^{-1} H(k,t) \mathcal{S} = -H(k, -t)$ with $\mathcal{S} = \sigma_z$.

The topological properties of the two-band Floquet model are characterized by two topological invariants, $W_0$ and $W_\pi$, which describe the topology associated with the 0-gap and $\pi$-gap, respectively. These are related to the winding numbers~ $\nu_0$ and $\nu_{T/2}$ of the Floquet-Bloch state $\left|\phi_{-}(k,0)\right\rangle $ and $\left|\phi_{-}(k,T/2)\right\rangle$ along the $k$-axis via (the quasi-energy of occupied bands satisfies $-\pi/T<\epsilon_-(k)<0$ ):
\begin{equation}
W_0 = \tfrac{1}{2}(\nu_0 + \nu_{T/2}), \quad W_{\pi} = \tfrac{1}{2}(\nu_{T/2}-\nu_0 ).
\end{equation}
Thus, any two configurations with distinct values of $W_0$ or $W_\pi$ must also differ in $\nu_0$ or $\nu_{T/2}$. A detailed proof is provided in the appendix~\cite{appendix}.
 This guarantees the existence of at least one spacetime point $(k,t)$ where the sum of the corresponding FFOs yields exactly one zero eigenvalue. As illustrated in Fig.~\ref{fig1}(b), the FFO configuration (top) where $W_0=1$ and $W_\pi=0$ corresponds to $\nu_0 = \nu_{T/2} = 1$, while the configuration (bottom) where $W_0=0$ and $W_\pi=1$ corresponds to $\nu_0 = -1$ and $\nu_{T/2} = 1$. Therefore, their $\text{FFOs}$ should exhibit different winding numbers along the $k$ direction in $Q(k, 0)$, which will consequently lead to the appearance of opposite Bloch vectors .

\begin{figure}[b!]
	\centering
	\includegraphics[angle=0,width=9.2cm]{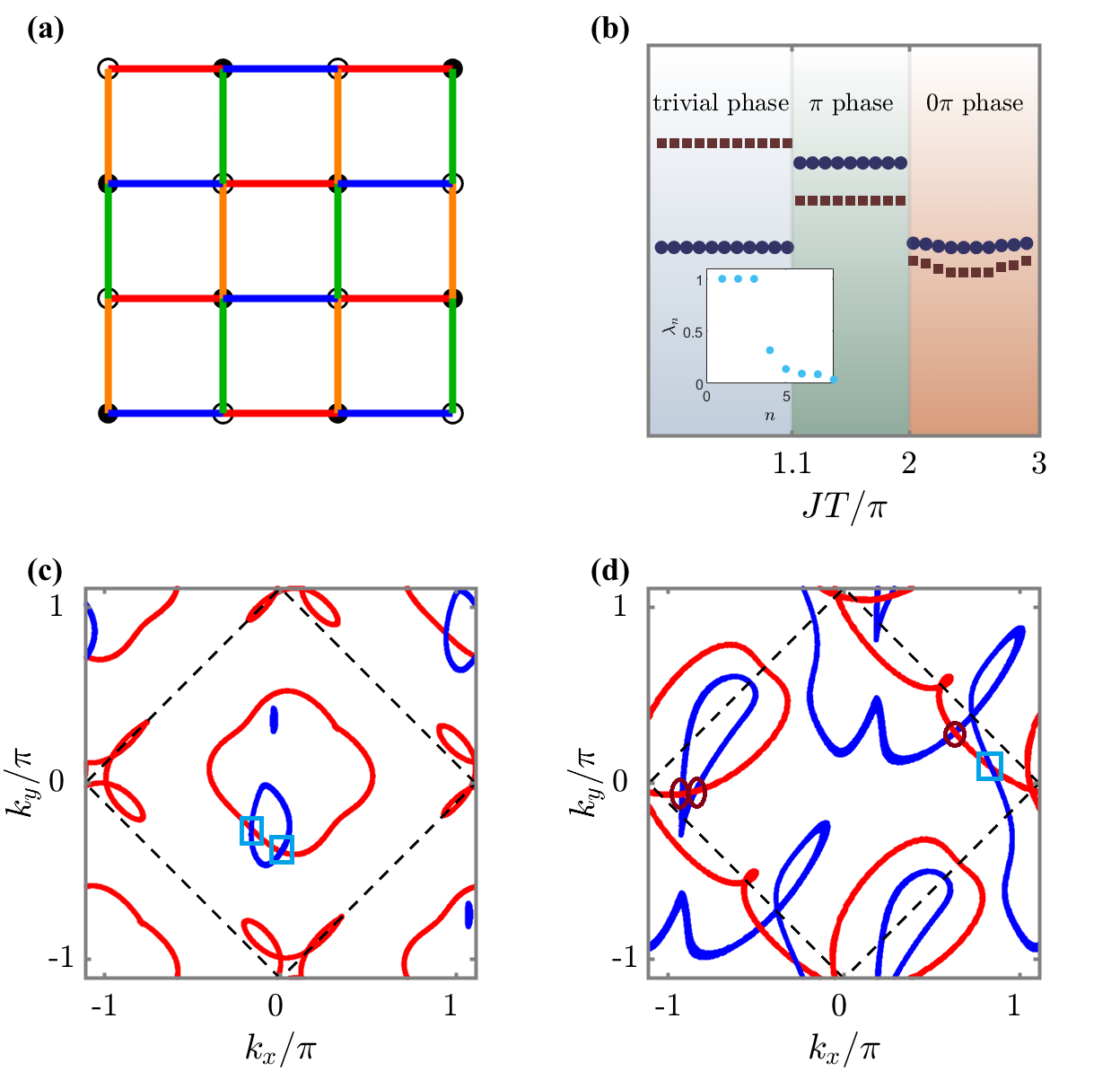}

	\caption{(a) Schematic illustration of the 2D modulated Floquet model, where different colors indicate the coupling between lattice sites in different time segments. The inter-atomic couplings evolve in a clockwise sequence following the order red, green, blue, and orange.
 (b) Phase diagram obtained using an unsupervised clustering algorithm. By comparing the clustering results with the theoretical topological invariants, three distinct phases are identified: the trivial phase with $W_0 = 0, \; W_\pi = 0$; the $\pi$ phase with $W_0 = 0, \; W_\pi = 1$; and the $0\pi$ phase with $W_0 = 1, \; W_\pi = 1$. (c,d) Inverse images of two Bloch vectors $\mathbf{n}$ and $-\mathbf{n}$ of the Floquet–Bloch states. The first Brillouin zone is indicated by the dashed lines; the linking number is 0 in (c) and 1 in (d).
(orientation of the Bloch vector $x=\cos(\theta)\sin(\phi)$, $y=\sin(\theta)\sin(\phi)$, $z=\cos(\phi)$, the red line: $\phi=\pi$, the blue line: $\theta=\pi,\phi=0.75\pi$.) (c) corresponds to the case with $J=0.4\pi/T$, $W_0 = 0, \; W_\pi = 0$, where the linking number of the preimages is 0; (d) corresponds to $J=2.7\pi/T$, $W_0 = 1, \; W_\pi = 1$, where the linking number of the preimages is 1.}
	\label{fig:fig2}
\end{figure}

We then constructed the dataset by sweeping the real and imaginary components of $\theta$ and analyzed it using our clustering algorithm. The spectral analysis of the diffusion map kernel, presented in Fig.~\ref{fig1}(d), reveals three dominant eigenvalues near unity, signifying the data is partitioned into three distinct clusters. Theoretical verification confirmed that these clusters map precisely to unique topological invariants, thereby demonstrating the algorithm's accuracy in identifying the topological phase diagram for this parameter region [Fig.~\ref{fig1}(c)].

\begin{figure}[b!]
	\centering
	\includegraphics[angle=0,width=9.2cm]{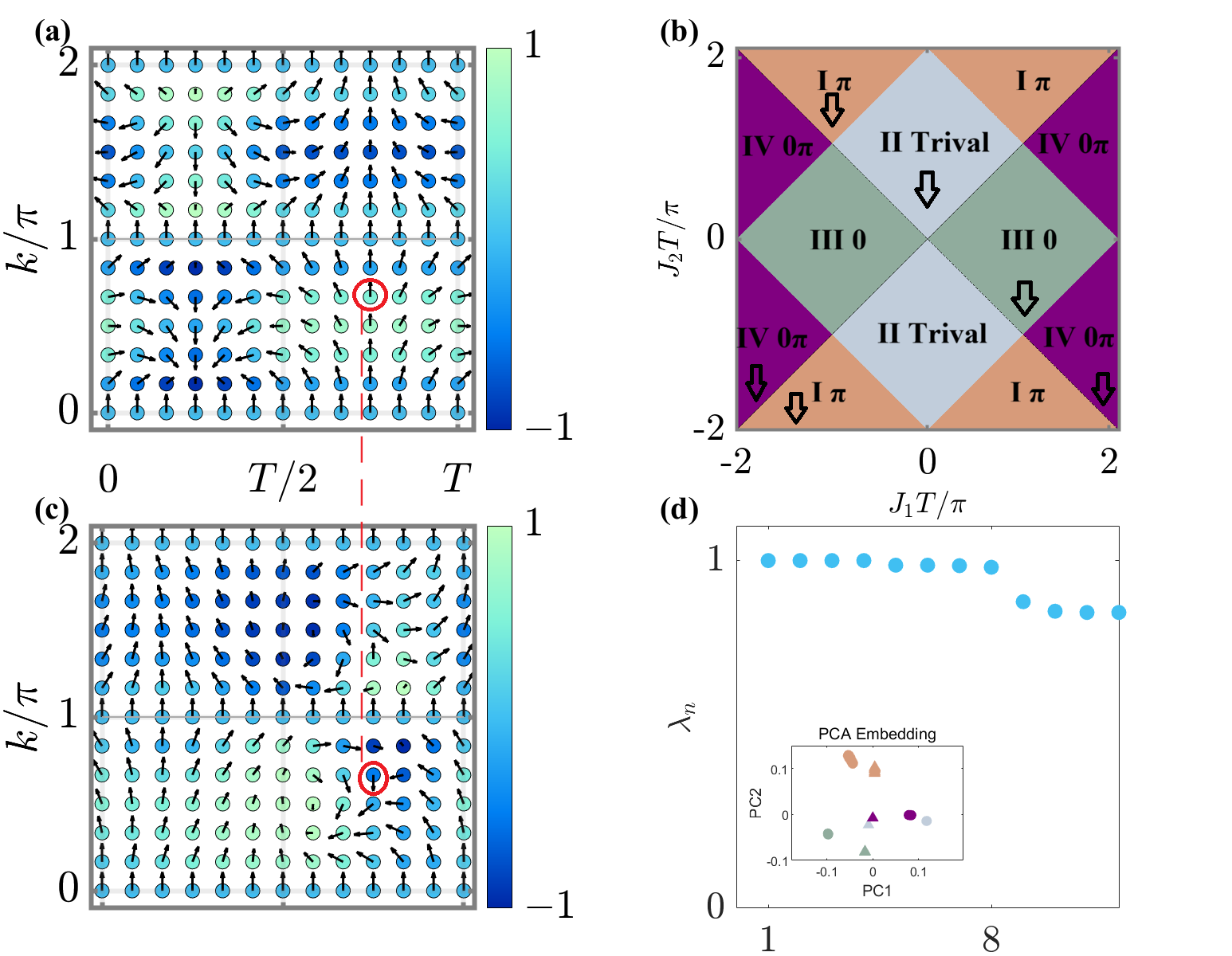}

	\caption{(a) (c)The Bloch vector representations for the FF0 for the modulated Floquet D model. The configuration in (a) is sampled from the Trivial-phase ($J_1=0,J_2=-\pi/T$) and the configuration in (c) is sampled from the $0\pi$-phase ($J_1=1.5\pi/T,J_2=\pi/T$). The colorbar describes the $z$-component of the Bloch vector.The red circles mark the flipping points identified by the FFO. (b) displays the phase diagram of the model, where the downward arrows indicate phases in which the Bloch vector at the high-symmetry point $k = 0$ point along the negative $y$ direction. (d) The eigenvalues output by the diffusion map algorithm, where eight eigenvalues close to 1 indicate that the data has been partitioned into eight clusters. The inset shows the two-dimensional projection of the dataset in the principal component space after sufficient diffusion. In the inset the triangles mark the phases in which the Bloch vector at the high-symmetry point $k = 0$ points along the negative $y$ direction. }
	\label{fig:fig3}
\end{figure}

\textit{A class in two-dimension}- To highlight the generalization capability of our approach, we examine the two-dimensional A-class~\cite{floqueta,floquetsingularities,floquetten}. Specifically, we focus on the following famous Floquet Bloch Hamiltonian:
\begin{equation}
H(\mathbf{k},t) =
\begin{cases}
J\sigma_{x}+\delta\sigma_z, & 0 \leqslant t < \frac{T}{5} \\
J\left(e^{i (k_{x}-k_{y})}\sigma^{+} + \text{H.c.}\right)+\delta\sigma_z, & \frac{T}{5} \leqslant t < \frac{2T}{5} \\
J\left(e^{i\left(2k_{x}\right)}\sigma^{+} + \text{H.c.}\right)+\delta\sigma_z, & \frac{2T}{5} \leqslant t < \frac{3T}{5} \\
J\left(e^{i (k_{x}+k_{y})}\sigma^{+} + \text{H.c.}\right)+\delta\sigma_z, & \frac{3T}{5} \leqslant t < \frac{4T}{5}\\
\delta\sigma_z, & \frac{4T}{5} \leqslant t \leqslant T
\end{cases}\label{Aclass}
\end{equation}
In real space, the Hamiltonian describes a periodically driven process on a square lattice as shown in Fig.~\ref{fig:fig3}(a). The hopping with strength $J$, is sequentially activated along each bond. This occurs in a clockwise sequence, with each activation lasting $T/5$. For the final $T/5$ of the period, all coupling is switched off. Additionally, a staggered on-site potential, with strength $2\delta$, is continuously applied to the $A$ and $B$ sublattices. This model belongs to the two-dimensional A-class, since it lacks time-reversal ($\mathcal{T}$), particle-hole ($\mathcal{P}$), and chiral ($\mathcal{S}$) symmetries~\cite{floquetten}. Its topological properties are characterized by two independent topological invariants, $W_0$ and $W_{\pi}$, corresponding to the quasienergy gaps at $0$ and $\pi/T$, respectively.

To explore different topological phases of this model, we fix $\delta = 0.5 \pi/T$ and generate the FFO data for the occupied band ($-\pi/T<\epsilon(\mathbf{k})<0$) with different $J$ which is uniformly sampled from $0$ to $3\pi/T$. The output of the diffusion map algorithm is presented in Fig.~\ref{fig:fig3}(b). The presence of three dominant eigenvalues near unity signifies that the dataset is partitioned into three distinct clusters. Crucially, the clustering results exhibit excellent agreement with the three phases defined by $W_0$ and $W_{\pi}$ : the trivial phase ($W_0=0,W_{\pi}=0$), the $\pi$-phase ($W_0=0,W_{\pi}=1$) and the $0 \pi$-phase ($W_0=1,W_{\pi}=1$).

We then investigate how the Floquet-Bloch states encode the distinct topological phases. Similar to the case studied previously, there should exist two topological invariants of $\left|\phi_{-}(\mathbf{k},t)\right\rangle$ that form a bijection with $W_0,W_{\pi}$. 

The first invariant is the Chern number $C$ of  $\left|\phi_{-}(\mathbf{k},t)\right\rangle$ in the $(k_x, k_y)$ plane at fix $t$, which is given by the following relation~\cite{floqueta,floquetten,floquetsingularities}:
\begin{equation}
    C = W_{0}- W_{\pi}.
\end{equation}
Thus, the FFO corresponding to the $\pi$-phase data possesses a non-trivial Chern number, $C=-1$, for all $(k_x, k_y)$ planes. Conversely, the FFO for the trivial or $0\pi$-phase data yields $C=0$. Consequently,  the $\pi$-phase FFO data can distinguished from the other ones. 

The second invariant is the Pontryagin's $\nu$ invariant. When $C = 0$, $\nu$ reduces to the Hopf invariant~\cite{quenchhopf, quenchu}. The Hopf invariant is geometrically defined as the linking number of the trajectories formed by the pre-images of two fixed Bloch vectors in the three-dimensional $(k_x, k_y, t)$ parameter space. As shown in Fig.~\ref{fig:fig3}(c,d), we calculate the linking number by counting the number of times the red and blue trajectories cross each other, within the first Brillouin zone marked by the dashed lines, the red circle indicates that the red trajectory is above, and the blue box indicates that the blue trajectory is above. For trivial phase, two trajectories do not link with each other, while in $0 \pi$-phase the linking number is $1$. The difference in the Hopf invariant explains how the algorithm distinguish trivial phase  from $0 \pi$-phase. We find that for general case, the Pontryagin's invariant~\cite{quenchv,hopf} is given by~\cite{appendix}:
\begin{equation}
    \nu = W_{\pi}\mod{2(W_{\pi}-W_0)}.
\end{equation}

It is worth noting that this relation provides an elegant method to calculate the Pontryagin's invariant for quantum quench problems with topological non-trivial initial states, which becomes complicated by the gauge dependence of the Chern-Simons integral. The detail of Pontryagin's is provided in the appendix~\cite{appendix}.

\textit{D class in one-dimension} - The third model we study belongs to symmetry class D and is described by the Hamiltonian~\cite{floquetd3}:

\begin{equation}
H(k,t) = \begin{cases}
    -J_1[\sin(k)\sigma_x - \cos(k)\sigma_y] + g\sin(k)\sigma_z, & t \leqslant  \frac{T}{2} \\
    J_2\sigma_y + g\sin(k)\sigma_z, & \hspace{-2em} \frac{T}{2} < t\leqslant T  
\end{cases}
\end{equation}
where $J_{1}$ and $J_{2}$ represent the strengths of
nearest spin interactions. The parameter $g$ breaks both chiral and time-reversal symmetries. The system retains particle-hole symmetry~\cite{floquetd3,floquetten}, defined by  $\mathcal{C}^{-1} H(k,t) \mathcal{C} = -H^*(-k, t)$ with $\mathcal{C} = I$.

In contrast to the $\mathbb{Z}$-type invariants of the previous models, the one-dimensional D Class system is characterized by a $\mathbb{Z}_2$ invariant, indicating two states for each quasienergy gap: trivial ($V_{0/\pi}=+1$) or nontrivial ($V_{0/\pi}=-1$). In static 1D D Class systems, the topological invariant is determined solely by the orientations of the Bloch vectors at the high-symmetry points $k=0$ and $k=\pi$. For this Floquet system, we find the relation between the topology of Floquet-Bloch state and $V_{0/\pi}$ is given by~\cite{appendix}
\begin{equation}
 V_{0} V_{\pi}=2{ \left|\left\langle\phi_{-}(\pi,t)|\phi_{-}(0,t) \right\rangle\right|^2-1},  V_\pi = (-1)^{C_{h}},
\end{equation}
where $C_{h}=\frac{1}{\pi}\int_{0}^{T}\!\mathrm{d}t\int_{0}^{\pi}\!\mathrm{d}k\
\mathrm{Im}\!\left\langle \partial_{t}\phi_{-}(k,t)\,\middle|\,\partial_{k}\phi_{-}(k,t)\right\rangle$~\cite{quenchbdi}. This implies that the orientations of the Bloch vectors at the high-symmetry lines $k=0$ and $k=\pi$ only determine the product $V_{0}V_{\pi}$, whereas the parity of $C_{h}$ dictates the triviality or nontriviality of $V_{\pi}$. $C_h$ has been discussed in D-class adiabatic pumps~\cite{floquetd4}. As shown in Fig.~\ref{fig:fig3}(a,c), the FFO reliably distinguishes the trivial phase and the $0\pi$ phase via the distinct $C_{h}$ values, even when their high-symmetry point orientations are identical.

We consider a dataset uniformly sampled from the parameter region  $J_1T/\pi\in[-2,2]$ , $J_2T/\pi\in[-2,2]$ and $g=0.5\pi/T$. Our algorithm ultimately classifies the data into eight distinct groups, as indicated by the eight nearly unitary eigenvalues shown in Fig.~\ref{fig:fig3}(d). These eight clusters correspond to the four theoretical phases shown in Fig.~\ref{fig:fig3}(b). The splitting of a single topological phase into two distinct subclasses occurs because the algorithm identifies different high-symmetry point orientations at $k=0$ and $k=\pi$ as separate groups, even when they map to the same $(V_0, V_\pi)$ invariant pair. Such additional sub-classifications are often unavoidable in automated symmetry-protected clustering, similar phenomenon has been observed in Ref.~\cite{diffusionlongyang}.




\textit{Conclusions and Discussion}— In summary, we have proposed an unsupervised approach for classifying Floquet topological phases based on FFO. By validating the method in three representative models---the one-dimensional AIII class, the one-dimensional D class, and the two-dimensional A class---we demonstrate its universality. This method allows us to study the topological classification of Floquet systems without requiring any prior knowledge of topological invariants, and they further reveal essential differences between Floquet topological phases and their static counterparts.

We also identify a correlation between the topological properties of Floquet–Bloch states and those of the quasienergy gaps. Moreover, the relation between the topological invariants of Floquet–Bloch states and gap topological invariants deserves further investigation. For instance, in the two-dimensional A class, distinct gap topological invariants may arise even when the Floquet–Bloch states share the same topological phase~\cite{appendix}. Similarly, in the one-dimensional D class, while the parity of the half–Brillouin-zone Chern number determines $V_\pi$, it remains unknown whether two configurations with the same parity of $C_{h}$ are homotopically connected. This raises the broader question of whether the topological properties of Floquet gaps necessarily coincide with those of the corresponding Floquet–Bloch states.

\textit{Acknowledgments}—This work is supported by the National Natural Science Foundation of China(NSFC) under Grant Nos.12204352(CW) and the National Natural Science Foundation of China (Grants No. 12174288 and No.12274326) and the National Key R \& D Program of China(Grant No. 2021YFA1400602).



\bibliography{Bibliography} 

\end{document}


\setlength{\parskip}{0pt}
\title{\texorpdfstring{Supplemental Material accompanying: \\ \emph{Unsupervised Topological Phase Discovery in Periodically Driven Systems via Floquet-Bloch State}}{Supplemental Material accompanying: \emph{Unsupervised Topological Phase Discovery in Periodically Driven Systems via Floquet-Bloch State}}}

\author{Chen-Yang Wang}
\affiliation{School of Physics Science and Engineering, Tongji University, Shanghai 200092, China}

\author{Jing-Ping Xu}
\affiliation{School of Physics Science and Engineering, Tongji University, Shanghai 200092, China}

\author{Ce Wang}
\email{cewang@tongji.edu.cn}
\affiliation{School of Physics Science and Engineering, Tongji University, Shanghai 200092, China}

\author{Ya-Ping Yang}
\affiliation{Tongji University, Shanghai 200092, China}

\maketitle

In this supplementary material, we provide a brief overview of the essential concepts regarding Floquet topology used in the main text. We primarily focus on the relationship between Floquet topological invariants and the topological properties of Floquet-Bloch states.

\section{Floquet Topological invariant theory}\label{sec:sec5}

A Floquet system is defined by a Hamiltonian that is periodic in time, $H(t) = H(t+T)$, with period $T$ and driving frequency $\omega = 2\pi/T$. The \textbf{effective Hamiltonian} $H^{\mathrm{eff}}_{\varepsilon}(\mathbf{k})$ serves as a natural starting point for analyzing these systems, as it maps the stroboscopic time-evolution operator $U(\mathbf{k}, T)$ onto a static-like framework~\cite{yao2017floquetten,rudner2013floquet2d}:

\begin{equation}
H^{\mathrm{eff}}_{\varepsilon}(\mathbf{k}) = \frac{i}{T} \ln_{-\varepsilon} [U(\mathbf{k}, T)]
\label{eq:eff_hamiltonian}
\end{equation}

where $\mathbf{k}$ denotes the $d$-dimensional momentum, and the logarithm $\ln_{\varepsilon}(e^{i\phi}) = i\phi$ is restricted to the branch $\phi \in (\varepsilon - 2\pi, \varepsilon]$. This formulation is particularly useful because it allows us to leverage well-established tools from static topological insulators to study the \textbf{quasienergy} spectrum---the eigenvalues of $H^{\mathrm{eff}}_{\varepsilon}(\mathbf{k})$.

However, while the effective Hamiltonian captures the physics at discrete intervals of $T$, it provides an incomplete picture of the system's topology. Due to the \textbf{anomalous bulk-boundary correspondence}~\cite{rudner2013floquet2d}, robust edge modes may exist even when the topological invariants derived from $H^{\mathrm{eff}}_{\varepsilon}(\mathbf{k})$ are entirely trivial. Consequently, the effective Hamiltonian alone cannot fully characterize a Floquet system. A complete topological description necessitates accounting for the \textbf{full time-evolution operator} $U(\mathbf{k}, t)$ throughout the entire driving cycle ($0 \le t < T$), ensuring that the dynamical information ``hidden'' between stroboscopic snapshots is preserved.

We now consider the topological characterization of the time-evolution operator $U(\mathbf{k}, t)$. In general, $U(\mathbf{k}, t)$ is not periodic in time because the evolution over a full period, $U(\mathbf{k}, T)$, does not necessarily return to the identity. To define a well-behaved topological invariant, one must construct a periodic counterpart $U_{\varepsilon}(\mathbf{k}, t)$ that satisfies $U_{\varepsilon}(\mathbf{k}, T) = U_{\varepsilon}(\mathbf{k}, 0) = I$.

As detailed in Ref.~\cite{rudner2013floquet2d}, this is achieved by defining a continuous deformation (homotopy) $U(\mathbf{k}, t; \alpha)$ that interpolates between the physical evolution ($\alpha=0$) and a periodic evolution ($\alpha=1$):
\begin{equation}
U_{\varepsilon}(\mathbf{k}, t) = U(\mathbf{k}, t) e^{i H_{\varepsilon}^{\mathrm{eff}}(\mathbf{k}) t}
\label{eq:periodic_U}
\end{equation}
The significance of this construction lies in the fact that it defines a continuous path connecting the non-periodic $U(\mathbf{k}, T)$ to the identity $I$, while ensuring that the quasienergy gap at $\varepsilon_\alpha = (1-\alpha)(\varepsilon+\pi/T)-\pi/T$ remains open throughout the entire deformation. Consequently, the topological properties of the $\varepsilon$-gap are preserved, allowing the system to be fully characterized by the winding properties of the periodic operator $U_{\varepsilon}(\mathbf{k}, t)$.

 When considering the symmetry constraints, we focus on the tenfold-way classification~\cite{yao2017floquetten}. Let $\mathcal{C}$, $\mathcal{T}$, and $\mathcal{S}$ denote the particle-hole, time-reversal, and chiral symmetry operators, respectively. Particle-hole and time-reversal symmetries are anti-unitary, satisfying $\mathcal{C}\mathcal{C}^* = \pm 1$ and $\mathcal{T}\mathcal{T}^* = \pm 1$, while the chiral symmetry $\mathcal{S}$ is a unitary composite of the two. In Floquet systems, these symmetries impose specific constraints on the instantaneous Hamiltonian $H(\mathbf{k}, t)$:
\begin{equation}
\mathcal{C}^{-1}H(\mathbf{k},t)\mathcal{C}=-H^*(-\mathbf{k},t), \quad \mathcal{T}^{-1}H(\mathbf{k},t)\mathcal{T}=H^*(-\mathbf{k},-t), \quad \mathcal{S}^{-1}H(\mathbf{k},t)\mathcal{S}=-H(\mathbf{k},-t)
\end{equation}

These relations propagate through the dynamics, dictating the transformation properties of the full time-evolution operator $U(\mathbf{k}, t)$:
\begin{equation}
\mathcal{C}^{-1}U(\mathbf{k},t)\mathcal{C}=U^*(-\mathbf{k},t), \quad \mathcal{T}^{-1}U(\mathbf{k},t)\mathcal{T}=U^*(-\mathbf{k},-t), \quad \mathcal{S}^{-1}U(\mathbf{k},t)\mathcal{S}=U(\mathbf{k},-t)
\end{equation}

Consequently, the effective Hamiltonian $H^{\mathrm{eff}}_{\varepsilon}(\mathbf{k})$ and the periodic evolution operator $U_{\varepsilon}(\mathbf{k},t)$ must satisfy the following symmetry-derived identities:
\begin{equation}
\mathcal{C}^{-1} H^{\mathrm{eff}}_{\varepsilon}(\mathbf{k}) \mathcal{C} = -H^{\mathrm{eff} *}_{-\varepsilon}(-\mathbf{k}) + \omega, \quad \mathcal{T}^{-1} H^{\mathrm{eff}}_{\varepsilon}(\mathbf{k}) \mathcal{T} = H^{\mathrm{eff} *}_{\varepsilon}(-\mathbf{k}), \quad \mathcal{S}^{-1}H^{\mathrm{eff}}_{\varepsilon}(\mathbf{k})\mathcal{S}=-H^{\mathrm{eff}}_{-\varepsilon}(\mathbf{k}) + \omega
\end{equation}
and
\begin{equation}
\mathcal{C}^{-1}U_{\varepsilon}(\mathbf{k},t)\mathcal{C}=U_{-{\varepsilon}}^*(-\mathbf{k},t)e^{i\omega t}, \quad \mathcal{T}^{-1}U_{\varepsilon}(\mathbf{k},t)\mathcal{T}=U_{\varepsilon}^*(-\mathbf{k},-t), \quad \mathcal{S}^{-1}U_{\varepsilon}(\mathbf{k},t)\mathcal{S}=U_{-{\varepsilon}}(\mathbf{k},-t)e^{i\omega t}.
\label{eq:u_symmetry}
\end{equation}

In the following sections, we introduce the Floquet topological properties of the systems discussed in the main text, with a particular focus on their connection to the topological characteristics of Floquet-Bloch states.

\section{Two-Dimensional A Class}

For the two-band model in the two-dimensional class A discussed in the main text, the periodic evolution operator $U_{\varepsilon}(\mathbf{k},t)$ possesses no additional symmetry. It defines a topological mapping from the three-torus to the unitary group, $S^1\times S^1\times S^1 \to U(2)$, where the domain $S^1\times S^1\times S^1$ is formed by the periodic coordinates $(k_x, k_y, t)$. Such mappings are classified by a 3D winding number~\cite{rudner2013floquet2d}:

\begin{equation}
W[U_\varepsilon] = \frac{1}{8\pi^2}\int_{0}^{T} dt \int_{\text{BZ}} dk_x dk_y \, \mathrm{Tr} \left( U_\varepsilon^{-1} \partial_t U_\varepsilon \left[ U_\varepsilon^{-1} \partial_{k_x} U_\varepsilon, U_\varepsilon^{-1} \partial_{k_y} U_\varepsilon \right] \right)
\label{eq:winding_number}
\end{equation}
where $[A, B] = AB - BA$ is the commutator. This invariant $W[U_\varepsilon]$ fully characterizes the topological phases associated with the $\varepsilon$-gap in class A systems. In accordance with the bulk-boundary correspondence, it is equivalent to the number of protected edge modes localized within the corresponding quasi-energy gap. 

We denote the winding numbers associated with the quasienergy gaps at $\varepsilon = 0$ and $\pi$ as $W_0$ and $W_\pi$, respectively. The topological properties of the Floquet-Bloch state $|\phi_{-}(\mathbf{k},t)\rangle$ can be directly related to these invariants. Specifically, for a fixed time $t$, the Chern number $C_{-}$ of the lower band is given by:
\begin{equation}
    C_{-} = W_0 - W_\pi \label{eq:2D_eq1}
\end{equation}

This result is consistent with the general relation established in Ref.~\cite{yao2017floquetten}, which states that $W[U_{\varepsilon'}] - W[U_{\varepsilon}] = c(P_{\varepsilon,\varepsilon'})$. Here, $P_{\varepsilon,\varepsilon'} = \sum_{\varepsilon < \varepsilon_n(\mathbf{k}) < \varepsilon'} |\phi_{n}(\mathbf{k},0)\rangle \langle \phi_{n}(\mathbf{k},0)|$ represents the projection operator onto the bands within the quasienergy interval $(\varepsilon, \varepsilon')$, and $c(P)$ denotes the associated Chern number. By setting the gap boundaries to $\varepsilon = -\pi/T$ and $\varepsilon' = 0$, the summation reduces to the single lower band (index $n \to -$), thereby recovering $C_{-} = c(P_{-\pi/T, 0})$.

However, this invariant alone does not provide a complete description of the system's Floquet topological information. For instance, the Chern number $C_-$ cannot distinguish between the trivial case ($W_0 = W_\pi = 0$) and the non-trivial case ($W_0 = W_\pi = 1$); the latter corresponds to the well-known regime of \textbf{anomalous Floquet topological insulators}~\cite{rudner2013floquet2d}, which host robust edge states despite having zero Chern numbers for all bulk bands. 

In fact, for a two-band system, the Floquet-Bloch state $|\phi_{-}(\mathbf{k},t)\rangle$ defines a mapping from the three-torus to the Bloch sphere, $S^1\times S^1\times S^1 \to S^2$. The full topological characterization of this mapping is captured by a set of four invariants: the three Chern numbers $\{C_{xy}, C_{xt}, C_{yt}\}$ associated with the different planes of the $(k_x, k_y, t)$ parameter space, and the Pontryagin's number $\nu$~\cite{Chen2020quench,hopf1}. In the context of Floquet-Bloch state, $C_{xt}$ and $C_{yt}$ are always 0, thus the topological property is captured by $C_{xy}$ and $\nu$. These invariants must be fundamentally related to  $W_0$ and $W_\pi$. Given that we have already established $C_{xy} = C_{-} = W_0 - W_\pi$, the remaining task is to find the explicit dependence of $\nu$ on the invariants $W_0$ and $W_\pi$.

We can draw inspiration from previous studies of quench dynamics, where $\nu$ is assigned a clear physical interpretation. When an initial state with a trivial Chern number ($c_i = 0$) is quenched into a  Hamiltonian with a Chern number $c_f$, the invariant $\nu$ identifies as the Hopf invariant~\cite{Wang2017quench,floquetquench}, satisfying $\nu = c_f$. In this situation, $\nu$ corresponds to the linking number of the trajectories formed by the pre-images of two distinct Bloch vectors in the three-dimensional $(k_x, k_y, t)$ parameter space. Mathematically, this invariant can be evaluated through the integral of the Chern-Simons form. It can be further proven that, within this framework, the invariant satisfies $\nu = W_\pi$. When the initial state possesses a non-zero Chern number ($c_i \neq 0$), $\nu$ can no longer be identified as a standard Hopf invariant~\cite{Chen2020quench}. In this regime, the Chern-Simons integral becomes a gauge-dependent quantity, and the integral expression for $\nu$ becomes significantly more complex. By careful analysis, the invariant is determined by the relation $\nu = c_f - c_i \pmod{2c_i}$~\cite{Hu2020quench,Chen2020quench}. We further explore the relationship between $\nu,c_f$ ($c_i = C_{-} = W_0 - W_\pi$) and  $W_0,W_\pi$ through the following design:
\begin{equation}
    H(\mathbf{k},t) = \begin{cases} 
        (1+\delta)H_f(\mathbf{k}), & 0 \le t \le \frac{T}{1+\delta} \\
        -H_i(\mathbf{k}), & \frac{T}{1+\delta} < t \le T 
    \end{cases}
\end{equation}
where $H_f(\mathbf{k})$ and $H_i(\mathbf{k})$ possess Chern numbers $c_f$ and $c_i$, respectively. We normalize $H_f(\mathbf{k})$ such that $\mathrm{Tr}[H_f(\mathbf{k})]=0$ and $\mathrm{Tr}[H_f^2(\mathbf{k})]=2$, which constrains its eigenvalues to $\pm 1$. By setting $T = \pi$, we ensure a complete evolution cycle under $H_f$ such that $e^{-i H_f T} = -I$. The parameter limit $\delta \to 0$ introduces an infinitesimal duration of evolution under $H_i$; this serves to pin the Floquet-Bloch state $|\phi_{-}(\mathbf{k},0)\rangle$ to the eigenstate of $H_i$, ensuring the band carries the initial Chern number $c_i$. This setup establishes a correspondence between the Floquet-Bloch state $|\phi_{-}(\mathbf{k},t)\rangle$ and the quench dynamics wavefunction. We can then compute the Floquet invariants $W_0$ and $W_\pi$ for different values of $c_i$ and $c_f$, as summarized in the table below:
\begin{table*}[h!]
\centering
\caption{The relationship between  ($c_i, c_f$) and  ($W_0, W_\pi, \nu$).}
\label{tab:winding_results}
\renewcommand{\arraystretch}{1.5} 
\begin{ruledtabular}
\begin{tabular}{l | c c c c c c} 
\textbf{Initial Chern number} $c_i$ & 0 & 1 & 0 & 1 & $-1$ & $-1$ \\
\textbf{Final Chern number} $c_f$   & $-1$ & $-1$ & 1 & 0 & 0 & 1 \\
\hline 
\textbf{Winding number} $W_0$       & $-1$ & $-1$ & 1 & 0 & 0 & 1 \\
\textbf{Winding number} $W_\pi$     & $-1$ & $-2$ & 1 & $-1$ & 1 & 2 \\
\textbf{Pontryagin's invariant} $\nu$       & $-1$ & 0 & 1 & 1 & 1 & 0 \\
\end{tabular}
\end{ruledtabular}
\end{table*}

From the data in Table~\ref{tab:winding_results}, we extract the following fundamental relations:
\begin{equation}
    W_\pi = c_f - c_i, \quad W_0 = c_f.
\end{equation}
Consequently, the 3D invariant $\nu$ is related to the gap winding numbers via:
\begin{equation}
    \nu = W_\pi \pmod{2W_\pi - 2W_0}.
\end{equation}

\section{One-Dimensional AIII Class}

We now consider the topological classification of the evolution operator $U_\varepsilon(k,t)$ for the one-dimensional AIII class. In this symmetry class, the presence of chiral symmetry $\mathcal{S}$ imposes specific constraints on the operators $U_0(k,T/2)$ and $U_\pi(k,T/2)$ as:
\begin{equation}
\mathcal{S}^{-1}U_{0}(k,T/2)\mathcal{S}=-U_{{0}}(k,T/2), \qquad 
\mathcal{S}^{-1}U_{\pi}(k,T/2)\mathcal{S}=U_{{\pi}}(k,T/2).
\label{eq:chiral_cond}
\end{equation}

For the two-band SSH model considered in the main text, the chiral symmetry is represented by the operator $\mathcal{S} = \sigma_z$. This symmetry constrains the evolution operators at the high-symmetry time $t=T/2$. Specifically, for the $0$ and $\pi$ gaps, the operators take the following structured forms:
\begin{equation}
U_0(k,T/2) = \begin{pmatrix}
0 & u_0^+(k) \\
u_0^-(k) & 0
\end{pmatrix}, \qquad 
U_{\pi}(k,T/2) = \begin{pmatrix}
u_{\pi}^+(k) & 0 \\
0 & u_{\pi}^-(k)
\end{pmatrix}.
\label{eq:U_forms}
\end{equation}
The Floquet topological numbers associated with the $0$ and $\pi$ quasienergy gaps are then determined by the winding~\cite{yao2017floquetten} of the off-diagonal and diagonal components, respectively:
\begin{equation}
    W_0 = \frac{i}{2\pi}\int_{\text{BZ}} (u_0^+)^{-1}\partial_k u_0^+ \, dk, \quad W_\pi = \frac{i}{2\pi}\int_{\text{BZ}} (u_\pi^+)^{-1}\partial_k u_\pi^+ \, dk.
\end{equation}

It is noteworthy that the components of the evolution operators satisfy the following relations~\cite{yao2017floquetten}:
\begin{equation}
    \int_{\text{BZ}} (u_\varepsilon^+)^{-1}\partial_k u_\varepsilon^+ \, dk = - \int_{\text{BZ}} (u_\varepsilon^-)^{-1}\partial_k u_\varepsilon^- \, dk, \quad (\varepsilon = 0, \pi).
\end{equation}
This can be understood by considering the total winding number of the unitary operator $U_\varepsilon(k,t)$ at a fixed time $t$. Since $U_\varepsilon(k,t)$ is smoothly connected to the identity operator $U_\varepsilon(k,0) = I$, its total winding number must be zero for any $t$. Consequently, for $t = T/2$, we have:
\begin{equation}
    \int_{\text{BZ}} \mathrm{Tr}\left[ U_\varepsilon^{-1}(k,T/2) \partial_k U_\varepsilon(k,T/2) \right] dk = \int_{\text{BZ}} \left( (u_\varepsilon^+)^{-1}\partial_k u_\varepsilon^+ + (u_\varepsilon^-)^{-1}\partial_k u_\varepsilon^- \right) dk = 0,\quad (\varepsilon = 0, \pi).
\end{equation}

We now elucidate the connection between the winding numbers $W_0$ and $W_\pi$ and the topological properties inherent in the Floquet-Bloch states $|\phi_{\pm}(k,t)\rangle$. We achieve this objective by examining the direct correspondence between the evolution operator $U_\varepsilon(k,t)$ and the states $|\phi_{\pm}(k,t)\rangle$. Consider the following relations:
\begin{equation}
  |\phi_{\pm}(k,t)\rangle = U(k,t)e^{i\epsilon_\pm t} |\phi_{\pm}(k,0)\rangle, \quad U_\varepsilon(k,t) = U(k,t)e^{iH_{\varepsilon}^{\mathrm{eff}}(k)t}, \label{AIII_1}
\end{equation}
where $-\pi/T<\varepsilon_{-}(k)<0$ and $0<\varepsilon_{+}(k)<\pi/T$ are the quasi-energy. According to the definition of the effective Hamiltonian $H_{\varepsilon}^{\mathrm{eff}}(k)$ in Eq.~\eqref{eq:eff_hamiltonian}, it follows that $U_\varepsilon(k,t) |\phi_{\pm}(k,0)\rangle \propto |\phi_{\pm}(k,t)\rangle$. Specifically, these two objects are equivalent up to a phase factor arising from the choice of the branch cut in the definition of $H_{\varepsilon}^{\mathrm{eff}}(k)$. To be more specific,
\begin{subequations}
\begin{align}
    H_{0}^{\mathrm{eff}}(k)|\phi_{-}(k,0)\rangle &= \epsilon_{-}(k)|\phi_{-}(k,0)\rangle, & H_{0}^{\mathrm{eff}}(k)|\phi_{+}(k,0)\rangle &= (\epsilon_{+}(k) - \omega)|\phi_{+}(k,0)\rangle, \\
H_{\pi}^{\mathrm{eff}}(k)|\phi_{-}(k,0)\rangle &= (\epsilon_{-}(k) - \omega)|\phi_{-}(k,0)\rangle, & H_{\pi}^{\mathrm{eff}}(k)|\phi_{+}(k,0)\rangle &= (\epsilon_{+}(k)-\omega)|\phi_{+}(k,0)\rangle.
\end{align} \label{AIII_2}
\end{subequations}
By combining Eq.~\eqref{AIII_1} and Eq.~\eqref{AIII_2}, we obtain the representation of the evolution operators at the half-period $t=T/2$:
\begin{align}
 U_0(k, T/2) &= -|\phi_{+}(k, T/2)\rangle \langle\phi_{+}(k, 0)| + |\phi_{-}(k, T/2)\rangle \langle\phi_{-}(k, 0)|, \label{eq:U0_decomp} \\
 U_\pi(k, T/2) &= -|\phi_{+}(k, T/2)\rangle \langle\phi_{+}(k, 0)| - |\phi_{-}(k, T/2)\rangle \langle\phi_{-}(k, 0)|. \label{eq:Upi_decomp}
\end{align}

Notably, the chiral symmetry imposes a constraint on the effective Hamiltonian such that $\mathcal{S}^{-1}H^{\mathrm{eff}}_{\varepsilon}(k)\mathcal{S}=-H^{\mathrm{eff}}_{-\varepsilon}(k) + \omega$. This symmetry protection allows the Floquet-Bloch states at $t=0$ to be parameterized as:
\begin{equation}
    |\phi_{\pm}(k,0)\rangle = \frac{1}{\sqrt{2}} 
    \begin{pmatrix} 
         e^{i\theta_0(k)} \\ \pm1
    \end{pmatrix}, \label{eq:phi_init}
\end{equation}
where $\theta_0(k)$ denotes the relative phase between the two sublattices in the chiral basis. When the Floquet cycle is initialized at $t = T/2$, the system's chiral symmetry is also preserved. Consequently, the Floquet-Bloch states $|\phi_{\pm}(k,T/2)\rangle$ can be parameterized as:
\begin{equation}
    |\phi_{\pm}(k,T/2)\rangle = \frac{e^{i\chi_{\pm}(k)}}{\sqrt{2}} 
    \begin{pmatrix} 
         e^{i\theta_1(k)} \\ \pm1
    \end{pmatrix}, \label{eq:phi_half}
\end{equation}
where $\chi_{\pm}(k)$ is a $k$-dependent phase factor and $\theta_1(k)$ represents the relative sublattice phase at $t = T/2$.
By substituting the parameterized Floquet-Bloch states from Eqs.~\eqref{eq:phi_init} and \eqref{eq:phi_half} into the spectral representations in Eqs.~\eqref{eq:U0_decomp} and \eqref{eq:Upi_decomp}, and subsequently comparing these with the matrix forms in Eq.~\eqref{eq:U_forms}, we obtain the following relations:
\begin{equation}
    \chi_{-}(k) = \chi_{+}(k) = \frac{\theta_0(k) - \theta_1(k)}{2}, \quad u_0^{+}(k) = e^{-i \frac{\theta_0(k) + \theta_1(k)}{2}}, \quad u_\pi^{+}(k) = e^{i \frac{\theta_0(k) - \theta_1(k)}{2}}.
\label{eq:phase_relations}
\end{equation}
Finally, we have 
\begin{equation}
    W_0= \frac{1}{2}(\nu_0+\nu_{T/2}),\quad W_\pi=\frac{1}{2}(\nu_0-\nu_{T/2}),
\end{equation}
where $\nu_0 = \frac{1}{2\pi}\int_\text{BZ}\partial_k\theta_0(k)dk$ and $\nu_{T/2} = \frac{1}{2\pi}\int_\text{BZ}\partial_k\theta_1(k)dk$ represent the winding numbers of the Bloch vectors associated with $|\phi_{-}(k,0)\rangle$ and $|\phi_{-}(k,T/2)\rangle$, respectively, as they traverse the equatorial plane.

\section{One-Dimensional D Class}

One-dimensional D-class systems possess particle-hole symmetry. Although D-class systems violate particle-number conservation, their dynamics are governed by time-dependent Bogoliubov–de Gennes equations. The computational approach for time evolution in Floquet D-class systems is detailed in Ref.~\cite{Dclass}, We get
\begin{equation}
\mathcal{C}^{-1}U_{\varepsilon}(k,t)\mathcal{C}=U_{-{\varepsilon}}^*(-k,t)e^{i\omega t}, \quad
\mathcal{C}^{-1}H(k,t)\mathcal{C}=-H^*(-k,t)
\end{equation}

We can prove that in a one-dimensional two-band system with particle-hole symmetry, the Hamiltonian at high-symmetry points possesses only the $\sigma_y$ component, and the particle-hole operator satisfies $\mathcal{C} = I$, where $I$ is the identity operator.
\begin{align}
H(k,t)=a(k,t)I+b(k,t)\sigma_x+c(k,t)\sigma_y+d(k,t)\sigma_z,  \\
-H^*(-k,t)=-a(-k,t)I-b(-k,t)\sigma_x+c(-k,t)\sigma_y-d(-k,t)\sigma_z    
\end{align}

At high-symmetry points $k=0,\, k=\pi$, we get the coefficients for $I$ is zero. $\mathcal{C}^{-1} \sigma_y \mathcal{C} = \sigma_y$. In d class $\mathcal{C} \mathcal{C}^* = 1$. We can get $\mathcal{C}\neq\sigma_y$, $\mathcal{C} = I$, and the coefficients for $\sigma_x$ and $\sigma_z$ are both zero. The Hamiltonian at high-symmetry points possesses only the $\sigma_y$ component.

For a static one-dimensional D-class Hamiltonian, the matrix can be transformed into a skew-symmetric form $-\mathrm{i}H(\mathbf{0}),-\mathrm{i}H(\mathbf{\pi})$ at high-symmetry points $k=0,k=\pi$, allowing the topological invariant to be computed via this skew-symmetric matrix. The Pfaffian~\cite{topology} ($\operatorname{Pf}$) of the skew-symmetric matrix serves to characterize the topological invariant of the system. The Pfaffian of a \(2n \times 2n\) skew-symmetric matrix \(A\) is defined as:
\begin{equation}
\operatorname{Pf}(X) = \frac{1}{2^n n!} \sum_{\tau \in S_{2n}} (-1)^{|\tau|} X_{\tau(1)\tau(2)} \ldots X_{\tau(2n-1)\tau(2n)},
\end{equation}

where $\tau$ denotes a permutation in the symmetric group $S_{2n}$, $\operatorname{sgn}(\tau)$ denotes the sign (or parity) of the permutation $\tau$, which is $+1$ for an even permutation and $-1$ for an odd permutation, $X_{\tau(2i-1), \tau(2i)}$ represents the matrix element of $X$ at row $\tau(2i-1)$ and column $\tau(2i)$.

The topology invariants $V$~\cite{topology} for a static one-dimensional D-class Hamiltonian
is \textbf{$\mathbb{Z}_2$ topological invariant}, the invariant can be expressed as
\begin{equation}
 V=\operatorname{sgn}\big( \operatorname{Pf}[-\mathrm{i}H(0)] \operatorname{Pf}[-\mathrm{i}H(\mathbf{\pi})] \big)
 \end{equation}
$V$ is topologically nontrivial when its equal $-1$. In a one-dimensional two-band system, this topological invariant has a simple physical interpretation: it takes the value $-1$ when the Bloch vectors at the two high-symmetry points are opposite, and $1$ when they are identical.

However, in Floquet systems, we cannot compute the two topological invariants solely by this method. Instead, we can define the topological number via the effective Hamiltonian~\cite{qidian}. 

\begin{equation}
 V_{0} V_{\pi}=\operatorname{sgn}\big( \operatorname{Pf}[-\mathrm{i}H^{\mathrm{eff}}_{-\pi}(0)T] \operatorname{Pf}[-\mathrm{i}H^{\mathrm{eff}}_{-\pi}(\mathbf{\pi})T] \big)
 \label{eq:heff}
 \end{equation}
 The topological number of the effective Hamiltonian is given by the product of the topological invariants $V_0$ and $V_{\pi}$. Here, $V_0$ and $V_\pi$ represent the $\mathbb{Z}_2$ topological invariants for the $0$-gap and $\pi$-gap, respectively.
 
The theory for topological invariants in Floquet D-class systems differs from the static case, as it intrinsically depends on the time-evolution properties. A comprehensive framework, which involves the introduction of an additional synthetic momentum dimension, is presented in Ref.~\cite{yao2017floquetten}. A simplified method for computing the topological invariant is provided in Ref.~\cite{Dclass,Dclass1}. The $\mathbb{Z}_2$ topological invariants $V_0$ and $V_\pi$ are determined by the number of eigenvalue crossings of the evolution operators $U(0,t)$ and $U(\pi,t)$ through $+1$ and $-1$, respectively~\cite{Jiang2011floquet}.

Next, we discuss the connection between $V_0$ and $V_{\pi}$ and the topological invariants of the Floquet--Bloch states.

Eq.~\eqref{eq:heff} indicates that a topological invariant of the Floquet--Bloch states associated with $V_0$ and $V_{\pi}$ is given by the orientation at the high-symmetry points.
\begin{equation}
 V_{0} V_{\pi}=2{ \left|\left\langle\phi_{-}(\pi,t)|\phi_{-}(0,t) \right\rangle\right|^2-1}
\end{equation}
Another topological invariant is inspired by one-dimensional quench systems in D class.
In one-dimensional Floquet systems in symmetry class~D, the orientations of the Floquet--Bloch states at the high-symmetry points are time independent and are restricted to the positive or negative direction along the $y$ axis. These are referred to as time fixed points~\cite{quench}. An $S^2$ sphere can be defined between the two fixed points, which allows us to define a Chern number $C_{h}$ over half of the Brillouin zone.
 \begin{equation}
 C_{h}=\frac{1}{\pi}\int_{0}^{T}\!\mathrm{d}t\int_{0}^{\pi}\!\mathrm{d}k\
\mathrm{Im}\!\left\langle \partial_{t}\phi_{-}(k,t)\,\middle|\,\partial_{k}\phi_{-}(k,t)\right\rangle
\end{equation}
By exploiting the symmetries of the Floquet--Bloch states, we show that the two Chern numbers defined over half of the Brillouin zone are not independent; their sum vanishes.
From the symmetries of the effective Hamiltonian, one can derive the symmetries of the Floquet--Bloch states.
 \begin{equation}
 H^{\mathrm{eff}}_{-\pi}(k)|\phi_{-}(k,0)\rangle = (-H^{\mathrm{eff} *}_{\pi}(-k) + \omega)|\phi_{+}^*(-k,0)\rangle=\varepsilon_{-}(k)|\phi_{-}(k,0)\rangle=\varepsilon_{-}(k)|\phi_{+}^*(-k,0)\rangle
\end{equation}
 \begin{equation}
U_{-\pi}(k,t)|\phi_{-}(k,0)\rangle=U_{\pi}^*(-k,t)e^{i\omega t}|\phi_{+}^*(-k,0)\rangle
\end{equation}
Here we derived $|\phi_{-}(k,0)\rangle=|\phi_{+}^*(-k,t)\rangle$. We substitute this expression into the equation 
\begin{equation}
\langle\phi_{-}(k, 0)|U^{-1}_{-\pi}(k,t)\sigma_a U_{-\pi}(k,t) |\phi_{-}(k,0)\rangle
\end{equation}
Here, $a=x,y,z$. By taking the complex conjugate of the above expression, we obtain the symmetry of the Bloch vector $({n}_x,{n}_y,{n}_z)$.
\begin{equation}
{n}_x(k,t)=-{n}_x(-k,t), \quad {n}_y(k,t)={n}_y(-k,t), \quad {n}_z(k,t)=-{n}_z(-k,t).
\end{equation}
It is then evident that, under this symmetry, the sum of the two Chern numbers defined over half of the Brillouin zone vanishes.
 \begin{equation}
 0=\frac{1}{\pi}\int_{0}^{T}\!\mathrm{d}t\int_{0}^{\pi}\!\mathrm{d}k\
\mathrm{Im}\!\left\langle \partial_{t}\phi_{-}(k,t)\,\middle|\,\partial_{k}\phi_{-}(k,t)\right\rangle+\frac{1}{\pi}\int_{0}^{T}\!\mathrm{d}t\int_{\pi}^{2\pi}\!\mathrm{d}k\
\mathrm{Im}\!\left\langle \partial_{t}\phi_{-}(k,t)\,\middle|\,\partial_{k}\phi_{-}(k,t)\right\rangle
\end{equation}
We perform numerical calculations using the Hamiltonian model introduced in the main text.
\begin{equation}
H(k,t) = \begin{cases}
    -J_1[\sin(k)\sigma_x - \cos(k)\sigma_y] + g\sin(k)\sigma_z, & t \leqslant  \frac{T}{2} \\
    J_2\sigma_y + g\sin(k)\sigma_z, & \frac{T}{2} < t \leqslant T.
\end{cases}
\end{equation}
We can then compute the $C_h$, $V_0$ and $V_\pi$ for different values of $J_1$ and $J_2$, as summarized in the  Table~\ref{tab:Ch}:
\begin{table*}[h!]
\centering
\caption{The relationship between  ($C_h$) and  ($V_0, V_\pi$).}
\label{tab:Ch}
\renewcommand{\arraystretch}{1.5} 
\begin{ruledtabular}
\begin{tabular}{l | c c c c c c} 
\textbf{Hamiltonian parameters} $J_1$ & 2 & 1 & 3 & 3.5 & $3.5$ & \\
\textbf{Hamiltonian parameters} $J_2$   & $1$ & $0.5$ & 2 & 1 & 3  \\
\hline 
\textbf{${Z}_2$ topological invariants} $V_0$   & $-1$ & $-1$ & 1 & 1 & 1  \\
\textbf{${Z}_2$ topological invariants} $V_\pi$     & $-1$ & $1$ & -1 & $1$ & 1  \\
\textbf{ Chern number} $C_h$       & $-1$ & 0 & 1 & 2 & 2 \\
\end{tabular}
\end{ruledtabular}
\end{table*}

We found that the parity of the Chern number $C_{h}$ in the half Brillouin zone is only related to the $V_{\pi}$ topological number. When $V_{\pi}=1$,  $C_{h}$ are even numbers. else,  $C_{h}$ are odd numbers. This equation has been discussed in d-class adiabatic pumps~\cite{pump}.
\begin{equation}
  V_\pi = (-1)^{C_{h}}
\end{equation}